\documentclass[twocolumn,showpacs,preprintnumbers,amsmath,amssymb]{revtex4}

\usepackage{graphicx}
\usepackage{dcolumn}
\usepackage{bm}

\def\lsim{\mathrel{\rlap{\lower4pt\hbox{\hskip1pt$\sim$}}\raise1pt\hbox{$<$}}}                
\def\gsim{\mathrel{\rlap{\lower4pt\hbox{\hskip1pt$\sim$}}\raise1pt\hbox{$>$}}}  
\begin{document}


\title{The $k$-essence scalar field in the context of Supernova Ia Observations}

\author{Abhijit Bandyopadhyay$^1$} \email{abhi.vu@gmail.com}
\author{Debashis Gangopadhyay$^{2,3}$} 
\author{Arka Moulik$^1$} 
 \affiliation{$^1$Department of Physics, Ramakrishna Mission Vivekananda University, Belur Math, Howrah 711202, India\\
$^2$S. N. Bose National Centre for Basic Sciences, Salt Lake, Kolkata 700 098, India\\
$^3$Present address: Department of Physics, Ramakrishna Mission Vivekananda University, Belur Math, Howrah 711202, India.}


\begin{abstract}
A $k$-essence scalar field model having (non canonical) Lagrangian of the form
$L=-V(\phi)F(X)$ where $X=\frac{1}{2}g^{\mu\nu}\nabla_{\mu}\phi\nabla_{\nu}\phi$ with constant $V(\phi)$ is 
shown to be consistent with  luminosity distance-redshift data
observed for type Ia Supernova.
For constant $V(\phi)$, $F(X)$ satisfies a scaling relation which is used to 
set up a differential equation involving the Hubble parameter $H$,
the scale factor $a$ and the $k$-essence field $\phi$.
$H$ and $a$ are extracted from SNe Ia data and using the differential 
equation the time dependence of the field $\phi$ is found to be:  
$\phi(t) \sim \lambda_0 +  \lambda_1 t +  \lambda_2 t^2$. 
The constants $\lambda_i$ have been determined. 
The time dependence is  similar
to that of the quintessence scalar field (having canonical 
kinetic energy) responsible for homogeneous inflation.
Furthermore,  the scaling relation and the obtained time dependence of the field
$\phi$ is used to determine the $X$-dependence of the function $F(X)$.
\end{abstract}

\pacs{95.36.+x,98.80.Cq,97.60.Bw.}

\maketitle

\section{Introduction}
\label{sec:1}
Measurement of luminosity distance of the type Ia 
Supernovae (SNe Ia) 
\cite{ref:Perlmutter,ref:Riess98,ref:Riess04,ref:Riess07,ref:Astier,ref:WoodVasey,ref:condata,ref:kowalski} 
during nearly last two  decades establishes that the universe
is presently undergoing a phase of accelerated expansion.
Observation of Baryon Acoustic Oscillations (BAO) 
\cite{ref:Eisenstein,ref:Cole,ref:Huetsi,ref:Percival}, Cosmic Microwave 
Background (CMB) radiations \cite{ref:Hinshaw,ref:komatsu}, 
power spectrum of matter  distributions in the universe 
provide other independent evidence in favour of this 
late-time cosmic acceleration. 
A general label for the source of this late-time cosmic acceleration is
``dark energy", which is a hypothetical 
unclustered form of energy  with 
negative pressure
- the negative pressure leading  to the 
cosmic acceleration by counteracting 
the gravitational collapse.
SNe Ia observations have  revealed that roughly 
70\% of the content of the 
present universe consists  of dark energy.

The very nature and origin of dark energy still 
remains a mystery despite many years of
research. There exist diverse theoretical approaches 
from different viewpoints aiming
construction of models for dark energy 
to explain the present cosmic acceleration.
These include the $\Lambda-CDM$ model \cite{ref:weinberg89} 
which fits well with the present cosmological 
observations  but is also plagued with the fine 
tuning problem from the standpoint of particle physics. 
Alternative field theoretic models based on 
consideration of specific forms of the
energy-momentum tensor with a negative 
pressure in Einstein's equation 
include quintessence \cite{ref:quint} and $k$-essence 
\cite{armen1,armen2,armen3,armen4,chiba,ark1,cal} 
where scalar fields with slowly varying potentials
and scalar field kinetic energy, respectively, 
drive the cosmic acceleration. 
There also exist other viable models of 
dark energy based on modification of geometric part of Einstein's
equation which include $f(R)$ gravity \cite{fr1,fr2,fr3}, 
scalar-tensor theories \cite{st1,st2,st3,st4,st5} and
brane world models \cite{brm1,brm2}.

In this work we try to address certain issues
related to the framework of $k$-essence scalar field model of dark energy
leading to interesting phenomenological consequences in the context of
SNe Ia observations. The $k$-essence models which involve actions with 
non-canonical kinetic terms are strong candidates for dark energy. 
A theory with a non-canonical kinetic term was first proposed by 
Born and Infeld in order to get rid of the infinite self-energy
of the electron~\cite{born}. Similar theories were also studied in 
~\cite{dirac}. Cosmology witnessed these models first in the 
context of scalar fields having non-canonical kinetic terms which drive 
inflation. Subsequently  $k$-essence models of dark matter and dark
energy were also constructed. Effective field theories arising from string theories also have non-canonical kinetic terms~\cite{callan,gib1,gib2,sen}.

The motivation for this work comes from the following.
A constant  potential in the $k$-essence (non-canonical) 
Lagrangian $L = F(X)V(\phi)$, where 
$X = \frac{1}{2}g^{\mu\nu}\nabla_\mu\phi \nabla_\nu\phi$,
ensures existence of a scaling relation 
$X\left(dF/dX\right)^2 = Ca^{-6}$, $C$ is some constant
\cite{scherrer,chimento1}.
Recently a Lagrangian for the  $k$-essence field has been set up
for the curvature constant $K=0$ in a homogeneous and isotropic universe
incorporating the above mentioned scaling relation
\cite{gango1,gango2}. 
Combining this formalism of non-canonical Lagrangian and
scaling relation with observational data
we determine ($i$) the time dependence of the 
spatially homogeneous scalar field $\phi(t)$ driving the late-time cosmic acceleration
and ($ii$) a form of
$F(X)$ satisfying the scaling relation. 

This is done as follows. 
For a spatially homogeneous scalar field $\phi(t)$, 
Einstein's field equations in presence of this scaling relation 
leads to the equation : $\dot{H} - K/a^2 - \dot{\phi}/a^3 = 0$ 
(derived below in Sec.\ \ref{sec:2}, Eq.\ \ref{eq:diffeq})
$K$ being the curvature constant.
This equation connects the time derivative of the field with the scale
factor ($a$) and Hubble parameter ($H$) associated with the expansion of
FLRW universe. Such an equation holds for a wide class 
of forms for $F(X)$ respecting the scaling relation.
The late-time temporal behaviour of the 
phenomenological values of the
parameters $H$ and $a$, on the other hand, 
can be extracted from the 
measured values of the luminosity distance of observed SNe Ia events with 
different redshift ($z$) values ranging from $0 < z \lsim 1.8$.
Use of these phenomenological values $H(t)$ and $a(t)$ in the 
above equation thus opens up a possibility of determining 
from  observational data the time dependence of the scalar field $\phi(t)$ 
in the  $k$-essence model with constant potential,
without prior knowledge of any specific 
form for $F(X)$.
In presence of the scaling relation
the temporal behaviour of the $k$-essence scalar field as obtained 
from the analysis of SNe 
Ia data is found to be fitted 
with the form:
$\phi(\tau) = \phi(\tau=0) + \lambda_1 \tau + \lambda_2 \tau^2$
to an appreciable extent of statistical precision.
$\tau$ represents a dimensionless time parameter 
(defined in Sec.\ \ref{sec:3}), with 
$\tau=0$ corresponding to the present epoch.
The numerical values of the
constant coefficients $\lambda_1$ and $\lambda_2$ are estimated directly 
from the analysis of observational data for three different possible values 
of the curvature constant $K$. Interestingly the observational data is 
instrumental in not providing rooms for further higher order terms
of $\tau$ in the obtained time dependence of $\phi$.
The $k$-essence scalar field responsible for late-time cosmic acceleration 
is found to have similar time evolution properties to that of a quintessence
field responsible for a homogeneous inflation.
It should be mentioned that there also    
exist  scenarios to unify inflation with late-time acceleration in certain 
interesting versions of modified gravity theories \cite{odin1,odin2,odin3}.

Using the scaling relation we can directly compute the
dependence of the function 
$F(X) - F(X_0)$ on $X$ from the observational data
upto the multiplicative constant (appearing in the scaling relation)
where $X_0$ is the value of $X$ at the present epoch.
For this, we first compute the dependence of 
$F(\tau)-F(\tau=0)$ on time $\tau$ by numerically integrating
the scaling relation. Then the time parameter can be
eliminated from the obtained time dependences 
of $X(=\frac{1}{2}\dot{\phi}^2)$ and $F(\tau) - F(\tau=0)$,
by numerically evaluating both quantities at same values
of time $\tau$. From this we subsequently obtain
the dependence of $F(X) - F(X=0)$ on $X$  along with $1\sigma$ 
and $2\sigma$ uncertainties owing to the uncertainties 
in the observational data. Note that, this construction of $F(X)$ 
in presence of the scaling relation,
directly from observation is very different 
from other constructions of $F(X)$ \cite{simon,barger,aasen}.

The paper is organised as follows. In Sec.\ \ref{sec:2} a brief review
is given of the $k$-essence scalar field model used to determine the time
dependence of the field $\phi$. Accordingly an equation is set up
to determine $\phi$ (Eq.\ (\ref{eq:diffeq}) below). 
In Sec.\ \ref{sec:3} we present the methodology
of analysis of SNe Ia data in order to determine the relevant parameters
to be subsequently used to find $\phi$ and $F(X)$ in Sec.\ \ref{sec:4}. 
 Discussions and conclusions are presented in Sec.\ \ref{sec:5}.

\section{The $k$-essence scalar field model and its contact with observation}
\label{sec:2}
We first indulge in a brief recollection \textit{i.e.}
\begin{eqnarray}
L &=& -V(\phi)F(X)=p  \label{eq:Lagrangian}\\
\rho &=& V(\phi) (F - 2XF_X) \label{eq:density} 
\end{eqnarray}
where $L$ is the  Lagrangian for the $k$-essence models,
$p$ is the pressure and $\rho$ is the energy density. $F(X)$ is a function of 
$X$  with $X={1\over 2}g^{\mu\nu}\nabla_{\mu}\phi\nabla_{\nu}\phi$, and 
$V(\phi)$ is a potential.
In a FLRW background the Einstein's field equations  give
\begin{eqnarray}
&& H^2 = \frac{8\pi G}{3}\rho - \frac{K}{a^2}\label{eq:ee00} \\ 
&& \dot H + 4\pi G(\rho + p) -K/a^{2} = 0 \label{eq:eeii} 
\end{eqnarray}
where $K$ is the curvature constant, $a$ is the scale factor and $H=\dot{a}/a$.
From these two equations we obtain the
$K$-independent continuity equation
\begin{eqnarray}
\dot\rho +3 H (\rho + p)&=& 0  \label{eq:continuity} 
\end{eqnarray} 
Using Eqs.\ (\ref{eq:Lagrangian}) and (\ref{eq:density}) in Eq.\  
(\ref{eq:continuity}) and taking the field $\phi$ to be homogeneous 
(so that $\nabla_i \phi = 0$ and  $X = \frac{1}{2}\dot{\phi}^2$)
we get the equation for the $k$-essence field as
\begin{eqnarray}
(F_X+2XF_{XX})\ddot{\phi}+3HF_X\dot{\phi}
+(2XF_X - F)\frac{V_\phi}{V} = 0
\label{eq:keom} 
\end{eqnarray}
where $V_\phi \equiv \frac{dV}{d\phi}$, $F_X \equiv \frac{dF}{dX}$ and $F_{XX} \equiv \frac{d^2F}{dX^2}$.

If we now choose the potential $V(\phi)$ to be a constant ($\phi$-independent),
then the third term in Eq.\ (\ref{eq:keom})
is absent, one obtains the scaling \cite{scherrer,chimento1}
\begin{eqnarray}
XF_X^2 &=& Ca^{-6} \label{eq:scaling} 
\end{eqnarray}
where $C$ is some constant. 
We stress here that this scaling relation reflects a fundamental aspect 
of $k$-essence model with constant potential $V$. The existence of a
scaling relation implies presence of relevant scales in the 
theory. Here this is realized for a constant potential. Throughout this 
work $V$ is a constant and the scaling relation is preserved.

Using Eqs.\ (\ref{eq:Lagrangian}), (\ref{eq:density}) and (\ref{eq:scaling}) 
we get from Eq.\ (\ref{eq:eeii})
\begin{eqnarray}
\dot{H} -\frac{K}{a^2} - 4\sqrt{2}\sqrt{C}\pi G V\frac{\dot{\phi}}{a^3}&=&0\,.
\end{eqnarray}
Rescaling the field as $4\sqrt{2}\sqrt{C}\pi G V \phi \to \phi$
the above equation can be rewritten as
\begin{eqnarray}
\frac{dH}{dt} - \frac{K}{a^2} - \frac{\dot\phi (t)}{a^3} &=& 0
\label{eq:diffeq}
\end{eqnarray}
Eq.\ (\ref{eq:diffeq}) is a differential equation incorporating $H$, $a$ and 
$\phi$. Exploiting this equation we can determine the time dependence of
the field $\phi$ from phenomenological values of $H$ and $a$ 
as extracted from SNe Ia data.
Here we note the following.
In Eq.\ (\ref{eq:keom}), 
if $V(\phi)$ is not a constant then to determine the time 
evolution of $\phi$ from this equation would require prior knowledge of 
specific forms for $F(X)$ and $V(\phi)$ as well. However, if $V(\phi)$ is a
constant (as we have considered), 
scaling relation (Eq.\ (\ref{eq:scaling})) 
ensures the existence of Eq.\ (\ref{eq:diffeq}).
Now determining the time evolution of $\phi$ from Eq.\ (\ref{eq:diffeq})
does not require prior knowledge of specific form of $F(X)$, 
knowing $H$ and $a$ is enough.

Moreover, using $X = \frac{1}{2}\dot{\phi}^2$, from the scaling relation (Eq.\ (\ref{eq:scaling})), it follows that
\begin{eqnarray}
\frac{dF}{dt} &=& 
\sqrt{2C}\frac{1}{a^3}\frac{d\dot{\phi}}{dt}\,.
\label{eq:dfdt}
\end{eqnarray}
Exploiting the time dependences of the scale factor $a$ and field $\phi$
as extracted from observational data 
we can numerically integrate Eq.\ (\ref{eq:dfdt}) to get
time dependence of the quantity $(F(t) - F(t=0))/\sqrt{2C}$.
Further, the time parameter can be eliminated from
the obtained time dependence of $(F(t) - F(t=0))/\sqrt{2C}$
and $X(t)$ by numerically 
evaluating both the quantities at same values of $t$
from which we subsequently obtain the $X$-dependence
the quantity $(F(X) - F(X_0))/\sqrt{2C}$.

 Now recall standard theories of homogeneous inflation 
\cite{ref:tsujikawa,ref:mukhanov}.
There the potential energy of a homogeneous scalar field 
$\phi$, usually called inflaton (described by Lagrangians with
canonical kinetic terms), leads to the exponential expansion
of the universe. The energy density and the pressure density of the 
inflaton field can be described respectively, as 
\begin{eqnarray}
\rho &=& \frac{1}{2}\dot{\phi}^2 + V(\phi) \label{eq11}\\
p &=& \frac{1}{2}\dot{\phi}^2 - V(\phi) \label{eq12}
\end{eqnarray}

Substituting these in  
Eqs.\ (\ref{eq:ee00}) (for $K=0$) and (\ref{eq:continuity}) we get 
$H^2 = (8\pi G/3)(\frac{1}{2}\dot{\phi}^2 + V(\phi))$ and 
$\ddot{\phi}+3H\dot{\phi}+ dV/d\phi = 0$. 
For a constant potential $V(\phi) (\gg \dot{\phi}^2)$,
$H$ is approximately a constant and the last equation implies
$\phi \sim A e^{-\gamma t} \sim \gamma_0 + \gamma_1 t + \gamma_2 t^2 + \cdots$. 
SNe Ia observations fitted to Eq.\ (\ref{eq:diffeq}) 
seem to indicate that a similar time evolution is also true for
$k$-essence scalar fields where the kinetic energy dominates over 
the potential energy and the Lagrangian has non-canonical kinetic
terms.

Here the following should be noted. It may be argued that if $V$ is 
a constant , all $k$-essence models behave as quintessence if 
$X={1\over2}\dot\phi ^{2}\to 0$ i.e. ${\cal L}=-VF(X)\sim -V[F(0) +X F'(0)]$.
However, there is a subtle difference here. From Eq.\ (\ref{eq11}) and
(\ref{eq12}), $p+\rho = \dot{\phi}^2 = 2X$ for quintessence whereas from
Eq.\ (\ref{eq:Lagrangian}) and (\ref{eq:density}), 
using scaling relation we get
$p+\rho = -2V\sqrt{C}a^{-3}\sqrt{X}$ for $k$-essence. Therefore for 
$X \neq 0$, $k$-essence cannot reduce to quintessence. For $X \to 0$,
$p+\rho$ for $k$-essence approaches zero faster than $p+\rho$ for 
quintessence.  Therefore use of the scaling relation prevents
this model for  $k$-essence from reducing to a model for quintessence for 
any $X$. It should be mentioned that for constant $V$ the scaling relation
is a fundamental attribute obtained from the hydrodynamic model and 
Einstein's equation. Therefore there is no chance of the quintessence and 
$k$-essence models coinciding for $X \to 0$. Moreover it will be shown 
below that $\dot{\phi}^2$ is non-zero from SNe Ia observations.

Here we mention another point.
As explained in \cite{ref:paddy}, since Friedmann equations relate $H^2(t)$ to
the total energy density of the universe, the best we can do from any geometrical
observation is to determine the total energy density of the universe at 
any given time. It is not possible to determine the energy densities  of
individual components from any geometrical observation.
Therefore observationally dark energy is difficult to pinpoint. Hence
distinguishing between quintessence and $k$-essence scalar fields is not 
possible by considering total energy densities. However, in the previous 
paragraph we have seen that for $\dot{\phi}^2 \to 0$ the theoretical behaviour of 
$p+\rho$ is different for quintessence and $k$-essence scenarios under certain 
conditions (scaling relation holds for $k$-essence only).

\section{Methodology of analysis of SNe Ia data}
\label{sec:3}

The SNe Ia data remain the key observational ingredient in determining 
cosmological parameters related to dark energy. 
One possible way of examining a theoretical model  of dark energy
involves parametrisation of quantities like luminosity distance or the
Hubble parameter or  the dark energy equation of state in terms of redshift.
The constants of such parametrisations
show up in the expressions for the 
cosmological quantities like the scale factor $a(t)$ 
and the Hubble parameter $H(t)$. The numerical values of these
constants can then be found by fitting it to the observational data.
In this work, we have assumed a closed form parametrisation of the 
luminosity distance, $d_L$  \cite{ref:paddy}
\begin{eqnarray}
d_L(\alpha,\beta,z) &=& \frac{c}{H_0} \left(\frac{z(1 + \alpha z)}{1+ \beta z}\right)
\label{eq:DL}
\end{eqnarray}

where $c$ is the speed of light and $H_0$ the value of the 
Hubble  parameter at the present epoch defined through the dimensionless
quantity $h$ by $H_0 = 100\ h\ $km s$^{-1}$ Mpc$^{-1}$.
The luminosity distance is related to the distance modulus
$\mu$ as
\begin{eqnarray}
\mu_{\rm th}(\alpha,\beta,z) 
&=& 5\log_{10}\Big{[}D_L(\alpha,\beta,z)\Big{]} + \mu_0 
\nonumber\\
&=&  5\log_{10}\left[ \frac{z(1+\alpha z)}{1+\beta z}\right] + \mu_0\,\,.
\end{eqnarray}
where 
\begin{eqnarray}
D_L(\alpha,\beta,z) & \equiv & 
\frac{H_0}{c}d_L(\alpha,\beta,z) = \frac{z(1 + \alpha z)}{1+ \beta z}
\label{eq:Dlab}
\end{eqnarray}
is called the Hubble free luminosity distance (dimensionless) and 
$\mu_0 = 42.38 - 5\log_{10} h$.

There exist different compilations of SNe Ia observations:
HST+SNLS+ESSENCE \cite{ref:Riess07,ref:WoodVasey,Davis},
SALT2 and MLCS data \cite{Kessler}, UNION \cite{ref:kowalski} and
UNION2 data \cite{Amanullah}.
These different groups tabulated the values of the distance modulus 
for different values of the redshift from the SNe Ia observations. 
The observed values of the distance modulus $\mu_{\rm obs}(z_i)$ 
corresponding to  measured redshifts $z_i$ are given in terms of the 
absolute magnitude $M$ and the apparent magnitudes $m_{\rm obs}(z_i)$ by 

\begin{eqnarray}
\mu_{obs}(z_i)=m_{obs}(z_i)-M\,.
\end{eqnarray} 
To obtain the best-fit values of the parameters $\alpha$ and $\beta$ from 
SNe Ia observations we perform a $\chi^2$ analysis which involves 
minimization of suitably chosen $\chi^2$ function with respect to the 
parameters $\alpha$ and $\beta$.

For our analysis of SNe Ia data we use the 
$\chi^2$ function considered in \cite{Xu},
where the methodology of the likelihood is  discussed in detail. 
The $\chi^2$ function for the 
analysis of SNe Ia data is first defined in terms of parameters $\alpha$, 
$\beta$ and $M^\prime\equiv \mu_0 +M$ (called the nuisance parameter) as
\begin{eqnarray}
&& \chi^2_{\rm SN}(\alpha,\beta,M^{\prime}) \nonumber\\
 &=& \sum_{i=1}^N\frac{(\mu_{obs}(z_i)-
\mu_{th}(\alpha,\beta,z))^2}{\sigma_i^2} \\
 &=& \sum_{i=1}^N\frac{(5\log_{10}(D_L(\alpha,\beta,z_i))-m_{obs}(z_i)+M^{\prime})^2}{\sigma_i^2} \quad
\end{eqnarray}
where $\sigma_i$ is the uncertainty in observed distance modulus and $N$ is the total number of data points. Its marginalisation over the nuisance parameter as $\bar{\chi}_{\rm SN}^2(\alpha,\beta) = \\     -2\ln\int_{-\infty}^{\infty} \exp\left[-\frac{1}{2} \chi^2(\alpha,\beta,M^{\prime})\right]dM^{\prime}$ leads to
\begin{eqnarray}
\bar{\chi}_{\rm SN}^2(\alpha,\beta) = P - (Q^2/R)+\ln (R/2\pi)
\end{eqnarray} 
with
\begin{eqnarray*}
\displaystyle P &=& \sum_{i=1}^N\frac{(5\log_{10}
(D_L(\alpha,\beta,z_i))-m_{obs}(z_i))^2
}{\sigma_i^2} \nonumber\\
\displaystyle Q  &=& \sum_{i=1}^N\frac{(5\log_{10}
(D_L(\alpha,\beta,z_i)) - m_{obs}(z_i))
}{\sigma_i^2} \nonumber\\
\displaystyle R &=&\sum_{i=1}^N\frac{1}{\sigma_i^2}
\end{eqnarray*}
The function $\chi^2(\alpha,\beta,M^\prime)$ has a minimum at $M^\prime=Q/R$ which gives the corresponding value of $h$ as $10^{(M-M^\prime+42.38)/5}$. Dropping the constant term $\ln (R/2\pi)$  from $\bar{\chi}^2_{\rm SN}$, the function
\begin{eqnarray}
\chi^2_{SN} (\alpha,\beta) &=& P - \frac{Q^2}{R}
\label{chifinal}
\end{eqnarray}
can be used in for the analysis.

Determination of Hubble parameter from observational measurements is another probe to the accelerated expansion of the universe attributed to the dark energy. Compilation of the observational data based on measurement of differential ages of the galaxies by Gemini Deep Deep Survey GDDS \cite{Abraham}, SPICES  and VDSS surveys provide the values of the Hubble parameter at 15  different redshift values \cite{simon,x39,x40,x41}. The $\chi^2$ function for the analysis of this observational Hubble data can be defined as

\begin{eqnarray}
\chi^2_{\rm OHD}(\alpha,\beta) &=& \sum_{i=1}^{15} \left [ 
\frac {H(\alpha,\beta;z_i) - H_{\rm obs}(z_i)} {\Sigma_i} \right ]^2 \,\,,
\end{eqnarray}
where $H_{\rm obs}$ is the observed Hubble parameter value at $z_i$ with uncertainty $\Sigma_i$.

Varying the parameters $\alpha$ and $\beta$ freely we minimize the $\chi^2$ function which is defined as 
\begin{eqnarray}
\chi^2(\alpha,\beta) &=& \chi^2_{\rm SN}(\alpha,\beta)  +  \chi^2_{\rm OHD}(\alpha,\beta)\,\,.
\end{eqnarray}

The values of the parameters $\alpha$ and $\beta$ at which minimum of 
$\chi^2$ is obtained are the best-fit values of these parameters for the 
combined analysis of the observational data from SNe Ia and OHD. We also 
find the 1$\sigma$ and 2$\sigma$ ranges  of the parameters $\alpha$ and 
$\beta$ from the analysis of the observational data discussed above. In this 
case of two parameter fit, the  1$\sigma$ (68.3\% confidence level) and 
2$\sigma$(95.4\% confidence level) allowed ranges of the parameters 
correspond to $\chi^2 \leq \chi^2_{\rm min} + \Delta\chi^2$, where 
$\Delta\chi^2=2.30(6.17)$ denotes the 1$\sigma$(2$\sigma$) spread in 
$\chi^2$ corresponding to two parameter fit. 

\begin{figure}
\begin{center}
\includegraphics[width=8.5cm, height=7cm]{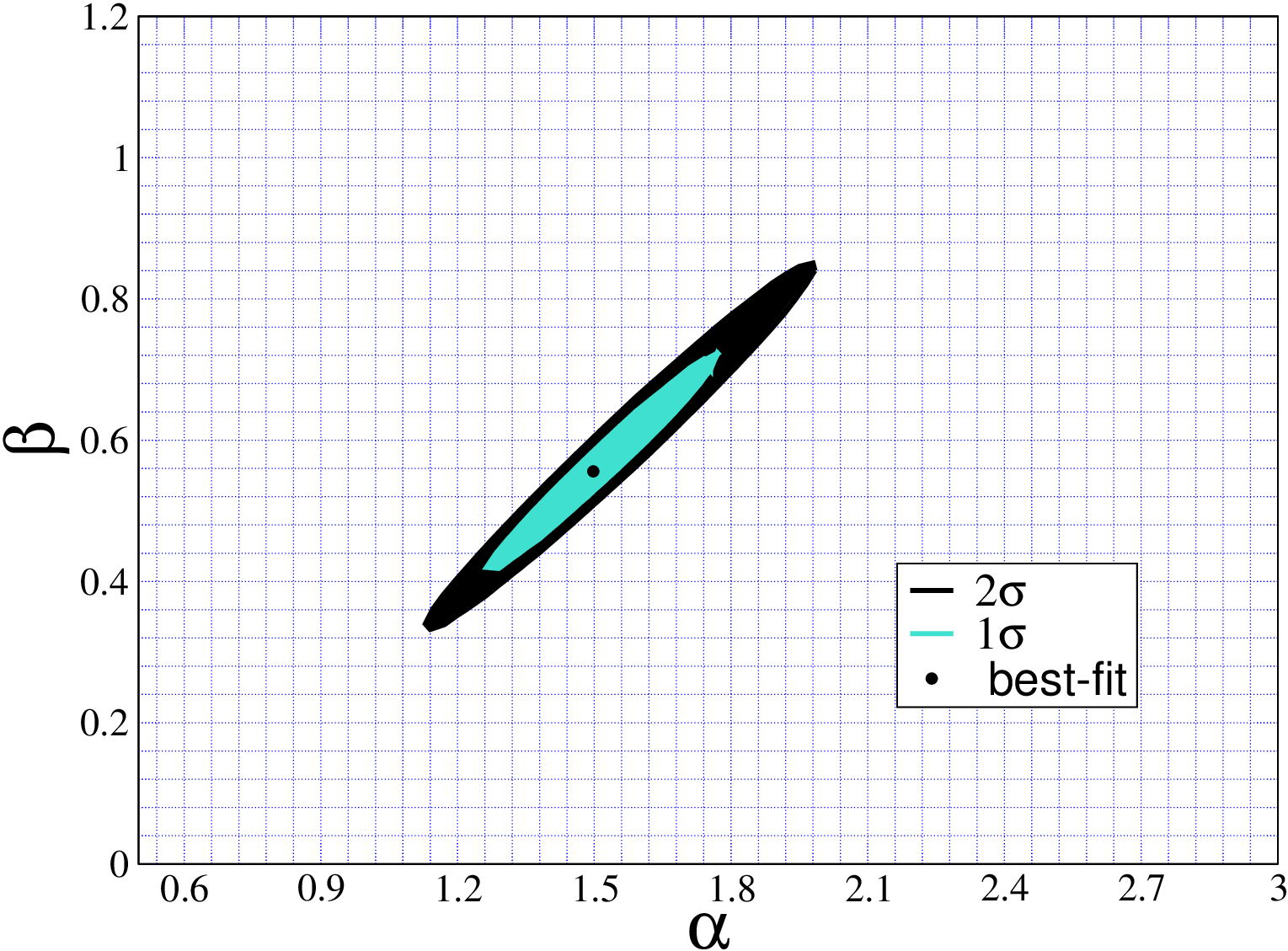}
\caption{Allowed regions in the parameter space $\alpha-\beta$ at 
1$\sigma$ and 2$\sigma$ confidence level from the combined analysis of SNe Ia data and OHD}
\label{fig:datafit}
\end{center}
\end{figure}

In this work we have considered the SNe Ia data from HST+SNLS+ESSENCE 
(192 data points)
\cite{ref:Riess07,ref:WoodVasey,Davis} and Observational Hubble Data from
\cite{simon,x39,x40,x41} (15 data points). 
The best fit for the combined analysis of the SNe Ia data and OHD 
is obtained for the parameters values
\begin{eqnarray}
\alpha =1.50 \quad , \quad \beta=0.55
\label{eq:ab}
\end{eqnarray}
with a minimum $\chi^2$ of 204.94. 
In Figure \ref{fig:datafit}
we have shown the regions of the $\alpha-\beta$ parameter space
allowed at $1\sigma$ and $2\sigma$ confidence levels from the analysis.
%
%
%

Using these values of $\alpha$ and $\beta$ as obtained from the
analysis we can determine the time dependence of the 
scale factor and the Hubble parameter.

\section{Determining $\phi(t)$ and $F(X)$ from the analysis of SNe Ia data}
\label{sec:4}
For a flat universe, which is consistent with
the current bounds from WMAP observation
on the ratio of the energy density in curvature to the critical density,
$|\Omega_K^{0}| < 0.01$ (95\% confidence level)\cite{ref:komatsu},
the Hubble parameter $H(z)$ corresponding to a redshift $z$ is directly
related to the luminosity distance through the relation
\begin{eqnarray}
E(z) 
&\equiv & \frac{H(z)}{H_0} 
 = \left[\frac{d}{dz}\left( \frac{D_L(z)}{1+z}\right)\right]^{-1}
\label{eq:Ez}
\end{eqnarray}
From the equations $H = \displaystyle\frac{\dot{a}}{a}$ and 
$\displaystyle\frac{a_0}{a} = 1 + z$
we get $dt = -\displaystyle\frac{dz}{(1+z)H}$.
Using this relation its useful to introduce a
dimensionless  parameter $\tau$ as
\begin{eqnarray}
\tau(z) &\equiv & H_0\Big{(}t(z) - t_0\Big{)} = - \int_0^z \frac{dx}
{(1+x)E(x)}\ .
\label{eq:tauz} 
\end{eqnarray}
The present epoch $(t(z=0) \equiv t_0)$ corresponds to $\tau(z=0) = 0$. In this work we consider the parameter $\tau$ to represent time and obtain the numerical solution for $d\phi/d\tau$ satisfying  
\begin{eqnarray}
\frac{d\phi}{d\tau} 
=    a^3(\tau) \frac{dE}{d\tau} - a(\tau) K \ ,
\label{eq:ftau}
\end{eqnarray}
as derived earlier in Eq.\ (\ref{eq:diffeq}).
%
\begin{figure}[t]
\begin{center}
\includegraphics[width=8.5cm, height=7cm]{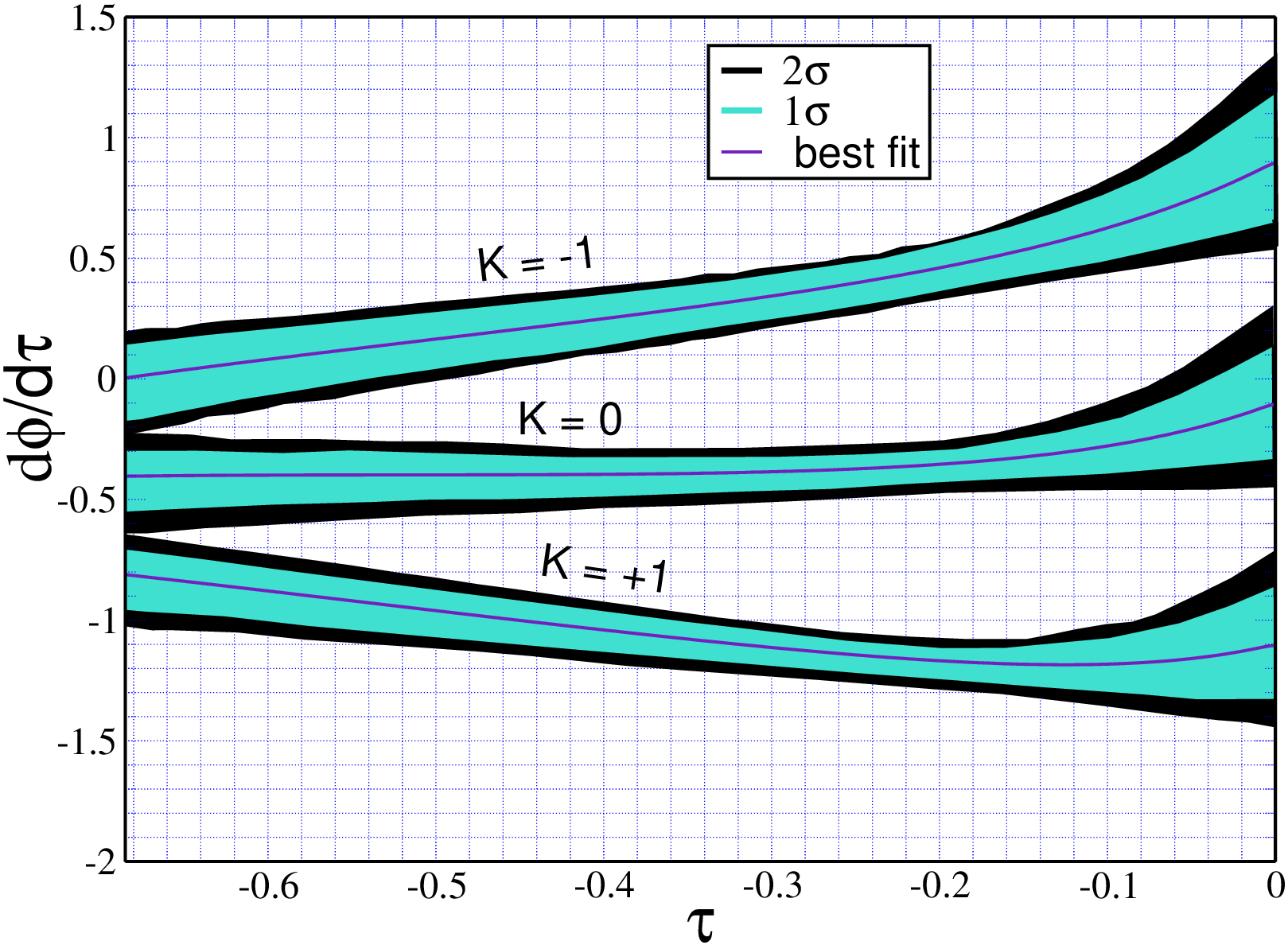}
\caption{Plot of $d\phi/d\tau$ vs $\tau$ for $K=+1, 0,-1$. Solid lines 
represent the plots drawn for best-fit values of $\alpha$ and $\beta$. The 
corresponding 1$\sigma$ and 2$\sigma$ variations are shown by shaded 
regions.}
\label{fig:ft}
\end{center}
\end{figure}
Using  Eqs.\ (\ref{eq:Dlab}), (\ref{eq:Ez}) and (\ref{eq:tauz}), with the values of $\alpha$ and $\beta$ as 
obtained from the analysis, we can numerically obtain the dependence of $E(z)$ and $\tau(z)$ on $z$. From these 
the time($\tau$)-dependence of $E$ can be extracted. We use this to evaluate $\frac{dE}{d\tau}$ as a function 
of $\tau$. Similarly from the equations $a=a_0/(1+z)$ and (\ref{eq:tauz}) the time variation of the scale 
factor can be obtained. All through the work we have assumed that the scale factor at present epoch is 
normalized to unity ($a_0=1$). Substituting the solutions for $dE/d\tau (\tau)$ and $a(\tau)$ in Eq.\ 
(\ref{eq:ftau}) we finally compute $d\phi/d\tau$ as a function of time ($\tau$). In Figure.\ \ref{fig:ft} we 
have shown the variation of $d\phi/d\tau$ with $\tau$ for all the three different values of the curvature 
constant $K$. The solid curves in this figure represent the estimated variation corresponding to the best-fit 
values of $\alpha$ and $\beta$ (Eq.\ (\ref{eq:ab}) ) obtained from the analysis.  Using the 1$\sigma$ and 
2$\sigma$ ranges of the parameters $\alpha$ and $\beta$ allowed from the analysis, we also obtain the 
corresponding spreads in the $d\phi/d\tau$-$\tau$ variation. These are shown by the shaded regions in Figure.\ 
\ref{fig:ft}.
It is important to note from the results that, for $K=0\ (+1)$,  $d\phi/d\tau 
= 0$ is disfavoured above 2$\sigma$ (1$\sigma$) level all of the epochs 
probed in SNe Ia observations. For $K=-1$, however, a zero value of 
$d\phi/d\tau$ remains allowed within 2$\sigma$ level  only for some earlier
epochs of time accessible in SNe Ia observations.

%
%

%
\begin{figure}
\begin{center}
\includegraphics[width=8.5cm, height=7.0cm]{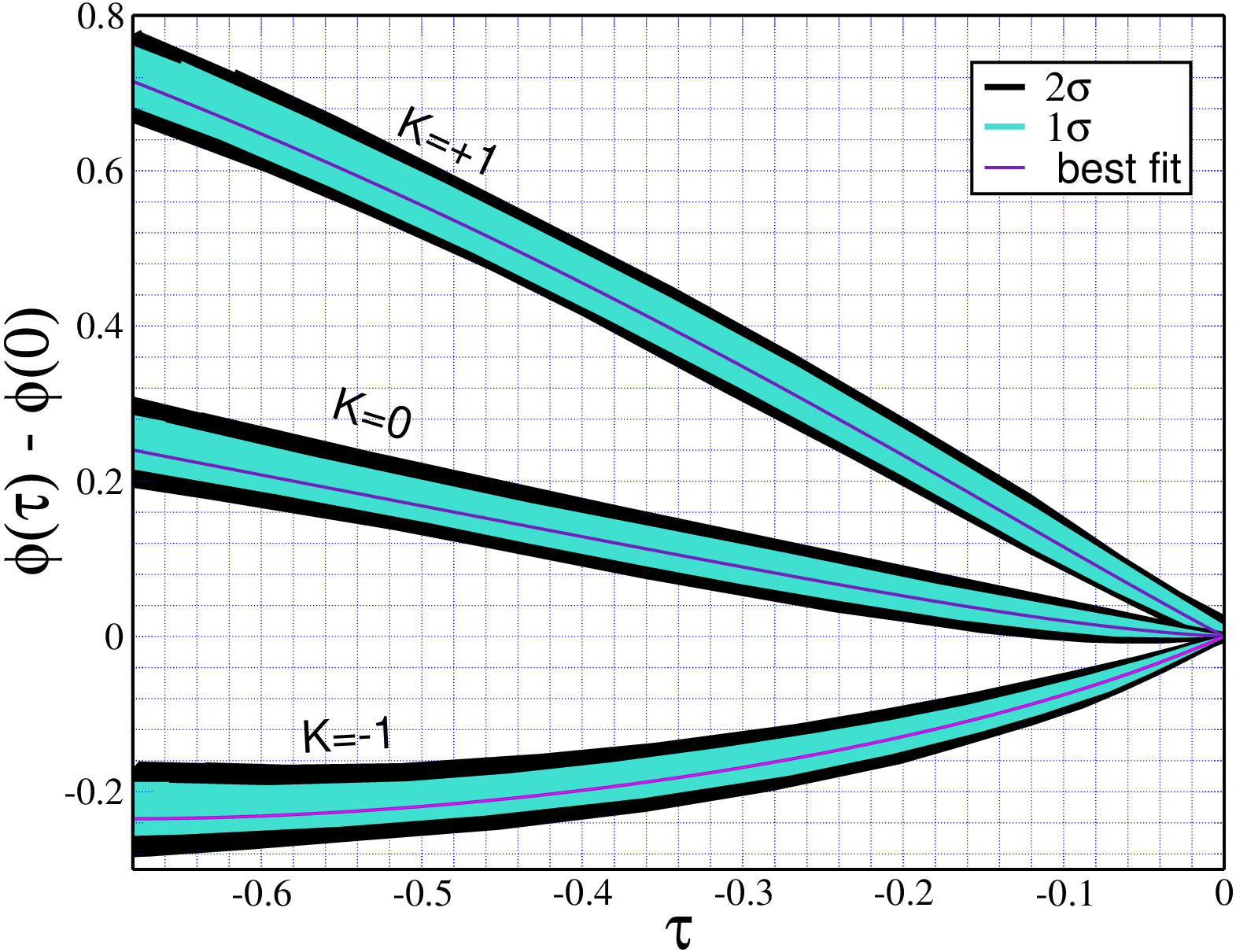}
\caption{Plot of $\phi(\tau) - \phi(0)$ vs $\tau$ for $K=+1, 0,-1$.
Solid lines represent the plots drawn for best-fit values of $\alpha$ and 
$\beta$. The corresponding 1$\sigma$ and 2$\sigma$ variations are shown by 
shaded regions.}
\label{fig:phi}
\end{center}
\end{figure}
\begin{figure}
\begin{center}
\includegraphics[width=8.5cm, height=7.0cm]{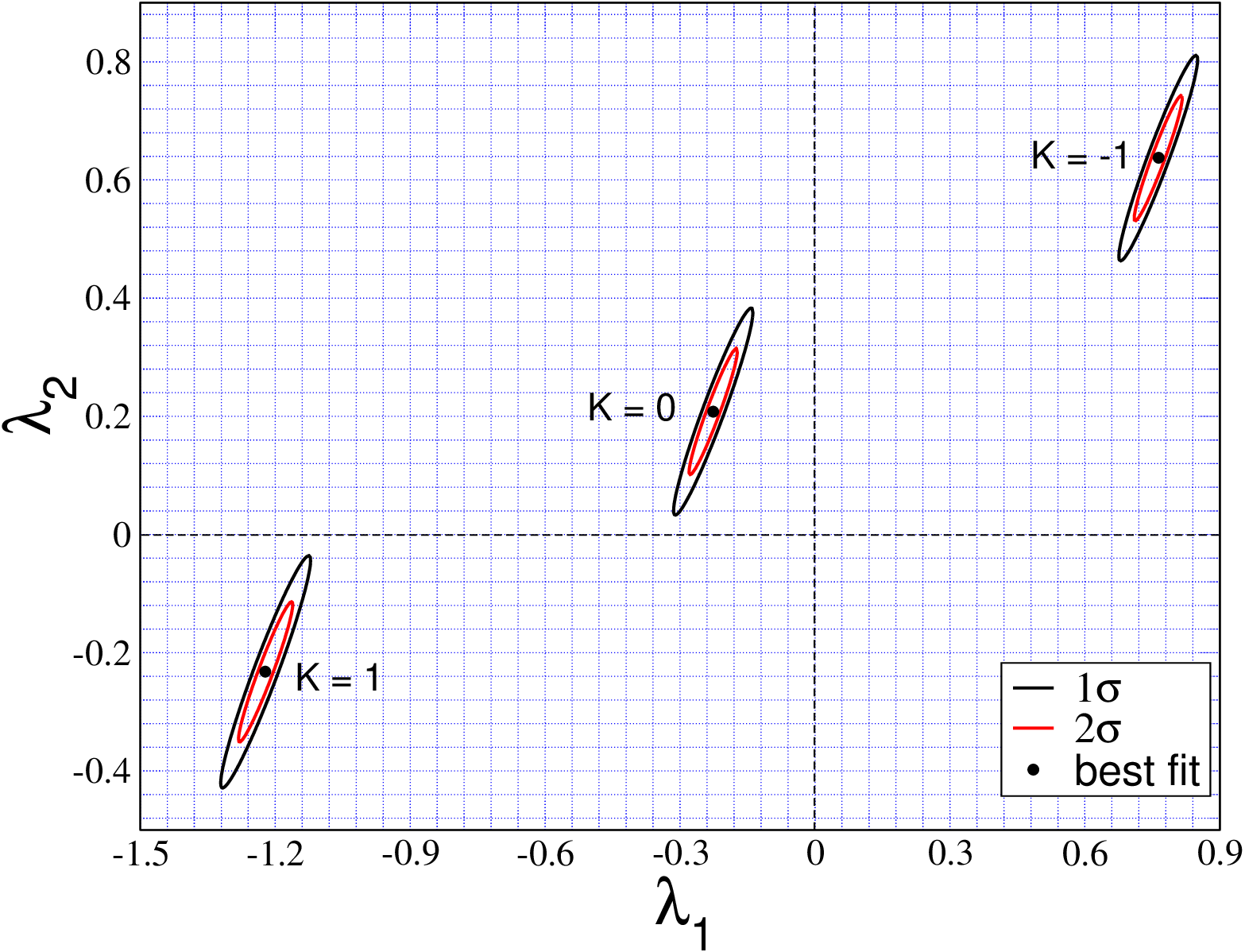}
\caption{Best-fit values of the parameters $\lambda_1$ and $\lambda_2$ with their 1$\sigma$ and 2$\sigma$ allowed ranges.}
\label{fig:l1l2}
\end{center}
\end{figure}

Integrating Eq.\ (\ref{eq:ftau}) and using 
the values of $\alpha$ and $\beta$ as obtained from the analysis,
we find that the obtained time dependence of 
$\phi(\tau)$ can  be fitted with 
\begin{eqnarray}
\phi(\tau) - \phi(0) & \sim & \lambda_1 \tau + \lambda_2 \tau^2
\label{eq:fitform}
\end{eqnarray}
to an appreciable extent of statistical precision. 
We present the results in Figure $\ref{fig:phi}$ 
where we have plotted $\phi(\tau) - \phi(0)$ 
as a function of  $\tau$ for three different values of $K$.  
The solid curves and shaded regions in Figure $\ref{fig:phi}$, respectively, 
represent the  plots corresponding to the best-fit values of  
$(\alpha ,\beta)$ and their  1$\sigma$ and 2$\sigma$ uncertainties.
The observational data are also instrumental in not providing room for further higher
order terms of $\tau$ in the obtained time dependence of $\phi$.
To find the values of $\lambda_1$ and $\lambda_2$ based on observational data and also to have a quantitative estimation of how good a form like Eq.\ (\ref{eq:fitform}) can accommodate the obtained time dependence of $\phi$ from analysis of observational data, we minimize the following $\tilde{\chi}^2$ function with respect to the parameters $\lambda_1$ and $\lambda_2$,
\begin{eqnarray}
\tilde{\chi}^2 
&=& \sum_{i=1}^n\left[\frac{\Phi_{\rm est}(\tau_i) - \Phi_{\rm fit}(\tau_i,\lambda_1,\lambda_2)}
{\Delta\Phi_{\rm est}(\tau_i)}\right]^2
\end{eqnarray}
where $\Phi_{\rm est}(\tau_i) \equiv \phi(\tau_i) - \phi(0)$ 
is estimated  in a way as described after Eq.\ (\ref{eq:ftau}) and 
$\Phi_{\rm fit}(\tau_i,\lambda_1,\lambda_2) = \lambda_1 \tau_i + 
\lambda_2 {\tau_i}^2$. $\Delta\Phi_{\rm est}(\tau_i)$ is the average value of 
estimated 1$\sigma$ uncertainty in $\Phi_{\rm est}(\tau_i)$. 
$n$ is the total number of points $(\tau_i,\Phi_{\rm est}(\tau_i))$
considered (distributed within time range accessible in SNe Ia observations)
for this particular analysis. We choose a large $n(=50)$ so that
$\tilde{\chi}^2_{\rm minimum}/n$ 
provides measure of $\tilde{\chi}^2$ per degree 
of freedom. The best-fit values of the parameters $\lambda_1$ and 
$\lambda_2$ from this analysis for all three different values of curvature 
constant along with the corresponding values of $\tilde{\chi}^2/n$ at the 
best-fit (minimum $\chi^2$) are listed in Table\ \ref{tab:1}. We also find 
the 1$\sigma$ and 2$\sigma$ allowed ranges for the parameters 
$\lambda_1$ and $\lambda_2$ which in this case of 2-parameter fit correspond
to $\tilde{\chi}^2 \leq \tilde{\chi}^2_{\rm minimum} + 2.30$ and 
$\tilde{\chi}^2 \leq \tilde{\chi}^2_{\rm minimum} + 6.17$, respectively. 
The ranges are shown by shaded regions in  Figure $\ref{fig:l1l2}$, where 
the best-fit points are also marked. 
At a quantitative level, very low values of $\tilde{\chi}^2_{\rm minimum}/n$
($\ll 1$) is a phenomenological indication of the fact that the time 
dependence of the 
$k$-essence field as extracted from observational data can be very 
convincingly accommodated within a profile: 
$\phi_{\rm k-essence}(t) \sim 
\lambda_0 + \lambda_1\ \tau + \lambda_2\ \tau^2$. The results of the analysis
show that for $K=1(-1)$ both $\lambda_1$ and $\lambda_2$ are negative
(positive) at 2$\sigma$ confidence level. For $K=0$, 
$\lambda_1$($\lambda_2$) is negative(positive) at 2$\sigma$ confidence 
level. 

\begin{figure}
\begin{center}
\includegraphics[width=8.5cm, height=7.0cm]{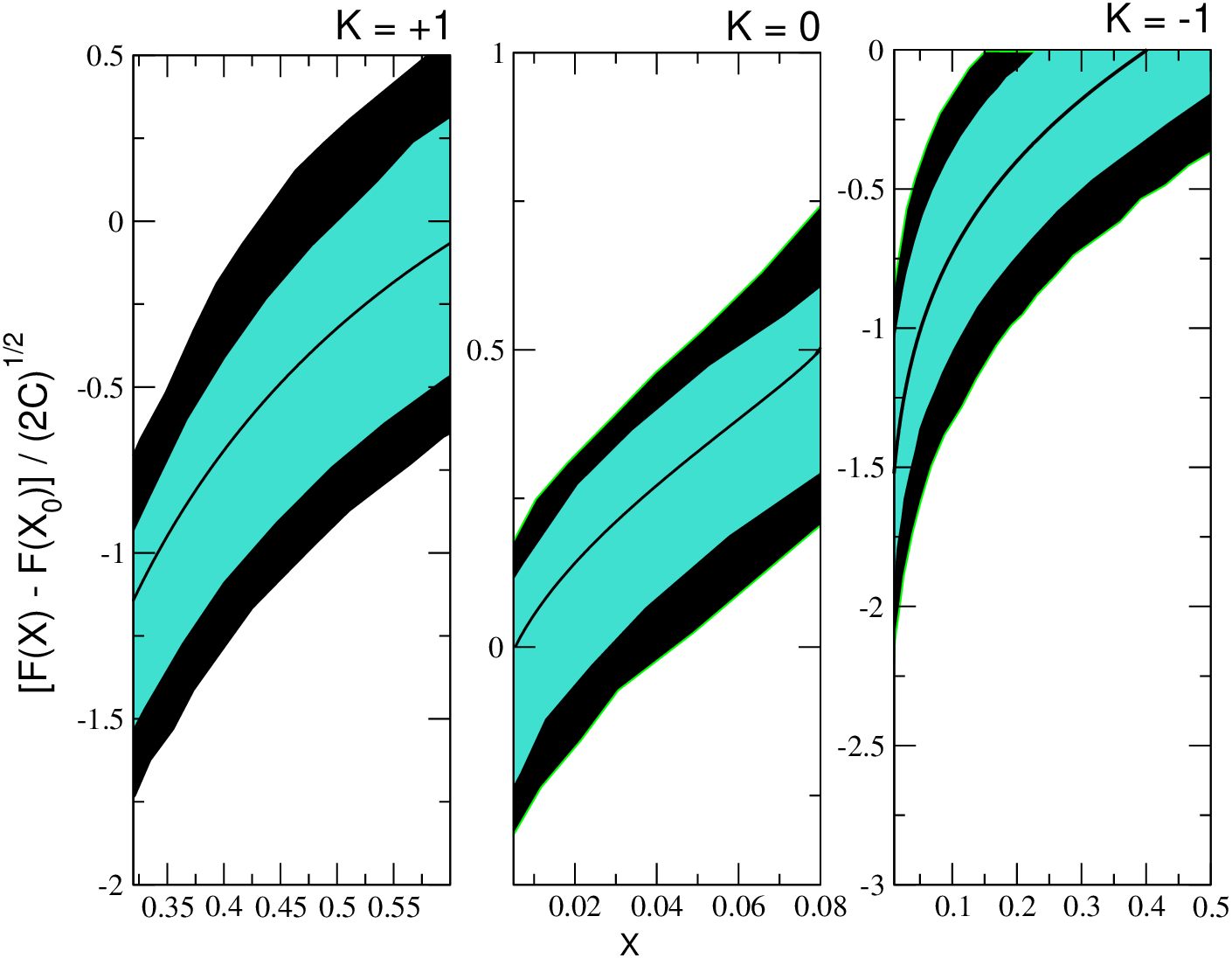}
\caption{Plot of the quantity $[F(X) - F(X_0)]/\sqrt{2C}$ vs $X$
with their 1$\sigma$ and 2$\sigma$ allowed ranges as obtained from analysis of observational data}
\label{fig:fxvsx}
\end{center}
\end{figure}

\begin{table}
\begin{center}
\caption{ Values of $\lambda_1$, $\lambda_2$ and $\tilde{\chi}^2_{\rm minimum}/n$ for $K=1,0,-1$.}
\label{tab:1}       
\begin{tabular}{cccc}
\hline\noalign{\smallskip}
$K$&  $\lambda_1$ &   $\lambda_2$ &   $\tilde{\chi}^2_{\rm minimum}/n$\\
\noalign{\smallskip}\hline\noalign{\smallskip}
+1 & -1.22  & -0.23 & 0.02\\
0 & -0.23 & 0.208 &  0.03\\
  -1 & 0.76 & 0.64 & 0.035\\
\noalign{\smallskip}\hline
\end{tabular}
\end{center}
\end{table}
%
%
Finally Eq.\ (\ref{eq:dfdt}) can be rewritten 
in terms of the dimensionless time parameter $\tau$.
Then using the $\tau$-dependence of the $k$-essence
scalar field $\phi$ as obtained above from the analysis of 
observational data, the $X$-dependence of the quantity 
$[F(X) - F(X_0)]/\sqrt{2C}$ can be found by the method as described in 
Sec.\ \ref{sec:2}. The plots are shown in 
Fig.\ \ref{fig:fxvsx} for three different values of curvature constant $K$.
This accommodates a wide spectrum of 
forms for the function $F(X)$ because of the arbitrariness 
of the constant $C$ appearing in the scaling relation. 
%

\section{Discussions and Conclusions}
\label{sec:5}
In this work we have shown that a model of $k$-essence
incorporating a  scaling relation 
can be accommodated within the luminosity distance - redshift data 
observed for type Ia Supernova. Existence of 
the scaling relation within the framework of $k$-essence model
allows to establish contact between the spatially 
homogeneous $k$-essence scalar field,
and the scale factor and the Hubble parameter associated
with the expansion of the FRW universe.
Estimation of the time dependences of  the 
cosmological quantities - the scale factor and
the Hubble parameter from the observed
SNe Ia data thus allows one to find temporal behaviour of  
the $k$-essence field $\phi(\tau)$. This comes out to be
$\phi(t) \sim \lambda_0 +  \lambda_1 \tau +  \lambda_2 \tau^2$  
where $ \lambda_i$ are constants and have been determined
from the analysis of the observational data along with their uncertainties.
With the obtained time dependence of the scalar field $\phi$,
the use of scaling relation helps to obtain the form of the 
function $F(X)-F(X_0)$ up to an arbitrary multiplicative 
constant occurring in the scaling relation itself. We have
also presented the 1$\sigma$ and 2$\sigma$ ranges
of this dependence as allowed from
the observational data. 

Time dependences of the $k$-essence scalar field
extracted from SNe Ia observations within its accessible 
time domain {\it viz} $0.9 <t/t_0<1.0$  
are similar to that of a quintessence field responsible for homogeneous
inflation which occurs at an epoch $t/t_0$ very close to zero. Note that
these two domains correspond to $-0.7<\tau<0$ ($k$-essence) 
and $\tau \to -\infty$ (quintessence),
respectively, in our treatment. The temporal domains of the
two scalar fields ($k$-essence and quintessence) 
are well separated.
Therefore, similar scalar fields 
(so far as their time evolution is concerned) 
can account for inflation as well as dark energy driven 
accelerated expansion.  Our work seems to indicate that this view 
has some observational support.
Also note that the scaling relation implies that for well 
behaved $F(X)$, $X$ cannot 
go to zero. This is borne out by the above results.

A more realistic model would have included dust matter. However,
this entails $\rho$ in Eqs.\ (\ref{eq:ee00}-\ref{eq:keom}) 
being replaced by 
$\rho + \rho_{\rm dust~matter}$. Further, a choice would have to be
made regarding the form of $\rho_{\rm dust~matter}$. Consequently
the scaling relation (Eq.\ \ref{eq:scaling}) is no longer valid. This problem is presently
under investigation.


\begin{thebibliography}{}


\bibitem{ref:Perlmutter}
  S.~Perlmutter {\it et al.},
  Astrophys.\ J.\  {\bf 517}, 565 (1999)

\bibitem{ref:Riess98}
  A.~G.~Riess {\it et al.},
  Astron.\ J.\  {\bf 116}, 1009 (1998)

\bibitem{ref:Riess04}
  A.~G.~Riess {\it et al.},
  Astrophys.\ J.\  {\bf 607}, 665 (2004)

\bibitem{ref:Riess07}
  A.~G.~Riess {\it et al.},
  Astrophys.\ J.\  {\bf 659}, 98 (2007)

\bibitem{ref:Astier}
  P.~Astier {\it et al.}
  Astron.\ Astrophys.\  {\bf 447}, 31 (2006)

\bibitem{ref:WoodVasey}
  W.~M.~Wood-Vasey {\it et al.},
  Astrophys.\ J.\  {\bf 666}, 694 (2007)

\bibitem{ref:condata}
  M.~Hicken {\it et al.},
  Astrophys.\ J.\  {\bf 700}, 1097 (2009)

\bibitem{ref:kowalski}
  M.~Kowalski {\it et al.}
  Astrophys.\ J.\  {\bf 686}, 749 (2008)


\bibitem{ref:Eisenstein}
  D.~J.~Eisenstein {\it et al.}
  Astrophys.\ J.\  {\bf 633}, 560 (2005)

\bibitem{ref:Cole}
  S.~Cole {\it et al.}
  Mon.\ Not.\ Roy.\ Astron.\ Soc.\  {\bf 362}, 505 (2005)

\bibitem{ref:Huetsi}
  G.~Huetsi,
  Astron.\ Astrophys.\  {\bf 449}, 891 (2006)

\bibitem{ref:Percival}
  W.~J.~Percival {\it et al.},
  Astrophys.\ J.\  {\bf 657}, 51 (2007)
  

\bibitem{ref:Hinshaw}
  G.~Hinshaw {\it et al.},
  Astrophys.\ J.\ Suppl.\  {\bf 180}, 225 (2009)

\bibitem{ref:komatsu}
  E.~Komatsu {\it et al.} ,
  Astrophys.\ J.\ Suppl.\  {\bf 192}, 18 (2011)


\bibitem{ref:weinberg89}
S.~Weinberg,  Rev.\ Mod.\ Phys.\ {\bf 61}, 1 (1989)




\bibitem{ref:quint}
Y.~ Fujii,  Phys.\ Rev.\ {\bf D26}, 2580 (1982);
R.~D.~Peccei, J.~Sola, and C.~Wetterich, Phys.\ Lett.\ \textbf{B195}, 183 (1987);
L.~ H.~ Ford, Phys.\ Rev.\ {\bf D35}, 2339 (1987);
C.~ Wetterich,  Nucl.\ Phys.\ {\bf B302}, 668 (1988);
B.~ Ratra and P.~ J.~ E.~ Peebles,  Phys.\ Rev.\ {\bf D37}, 3406 (1988);
Y.~ Fujii and T.~ Nishioka, Phys.\ Rev.\ {\bf D42}, 361 (1990);
T.~ Chiba, N.~ Sugiyama, and T.~ Nakamura, Mon.\ Not.\ Roy.\ Astron.\ Soc.\ 289 (1997);
P.~ G.~ Ferreira and M.~ Joyce, Phys.\  Rev.\  Lett.\ {\bf 79}, 4740 (1997);
P.~ G.~ Ferreira and M.~ Joyce, Phys.\ Rev.\ {\bf D58}, 023503 (1998);
R.~ R.~ Caldwell, R.~ Dave, and P.~ J.~ Steinhardt,  Phys.\ Rev.\ Lett.\ {\bf 80}, 1582 (1998);
S.~ M.~ Carroll,  Phys.\ Rev.\ Lett.\ {\bf 81}, 3067 (1998);
E.~ J.~ Copeland, A.~ R.~ Liddle, and D.~ Wands, Phys.\ Rev.\ {\bf D57}, 4686 (1998);
I.~ Zlatev, L.~ M.~ Wang, and P.~ J.~ Steinhardt, Phys.\ Rev.\ Lett.\ {\bf 82}, 896 (1999);
P.~ J.~ Steinhardt, L.~ M.~ Wang, and I.~ Zlatev, Phys.\ Rev.\ {\bf D59}, 123504 (1999);
A.~ Hebecker and C.~ Wetterich,  Phys.\ Rev.\ Lett.\ {\bf 85}. 3339 (2000);
A.~ Hebecker and C.~ Wetterich, Phys.\ Lett.\ {\bf B497}, 281 (2001)



\bibitem{armen1}
C.Armendariz-Picon, T.Damour and V.Mukhanov, Phys.Lett.
{\bf B458} 209 (1999).

\bibitem{armen2}
C.Armendariz-Picon, V.Mukhanov and P.J.Steinhardt, Phys.Rev.
{\bf D63} 103510 (2001).

\bibitem{armen3}
C.Armendariz-Picon, V.Mukhanov and P.J.Steinhardt,
Phys.Rev.Lett.
{\bf 85} 4438 (2000)

\bibitem{armen4}
C.Armendariz-Picon and E.A.Lim, JCAP {\bf 0508}(2005)
007

\bibitem{chiba}
T.~Chiba, T.~Okabe and M.~Yamaguchi, Phys.Rev.
{\bf D62} 023511 (2000)

\bibitem{ark1}
N.~Arkani-Hamed, H.~C.~Cheng, M.~A.~Luty and S.~Mukohyama, JHEP {\bf05} (2004) 074, JCAP {\bf0404} (2004) 001

\bibitem{cal}
R.R.Caldwell, Phys.Lett.{\bf B545} (2002) 23


\bibitem{fr1} S.~ Capozziello,  Int.\ J.\ Mod.\ Phys.\ {\bf D11}, 483  (2002)

\bibitem{fr2} S.~ Capozziello, V.~ F.~ Cardone, S.~ Carloni, and A.~ Troisi, Int.\ J.\ Mod.\ Phys.\ {\bf D12}, 1969 (2003)

\bibitem{fr3} S.~ M.~ Carroll, V.~ Duvvuri, M.~ Trodden, and M.~ S.~ Turner,  Phys.\ Rev.\ {\bf D70}, 043528 (2004)


\bibitem{st1} L.~ Amendola, Phys.\ Rev.\ {\bf D60}, 043501 (1999)
\bibitem{st2} J.~ P.~ Uzan,Phys.\ Rev.\ {\bf D59}, 123510 (1999)
\bibitem{st3} T.~ Chiba, Phys.\ Rev\. {\bf D60}, 083508, (1999)
\bibitem{st4} N.~ Bartolo and M.~ Pietroni, Phys.\ Rev.\ {\bf D61}, 023518 (2000)
\bibitem{st5} F.~ Perrotta, C.~ Baccigalupi, and S.~ Matarrese, Phys.\ Rev.\ {\bf D61}, 023507  (2000)


\bibitem{brm1} G.~ R.~ Dvali, G.~ Gabadadze, and M.~ Porrati, Phys.\ Lett.\ {\bf B485}, 208 (2000)
\bibitem{brm2} V.~ Sahni and Y.~ Shtanov,  JCAP {\bf 0311}, 014 (2003)

\bibitem{born}
M.Born and L.Infeld,  Proc.Roy.Soc.Lond {\bf A144}(1934) 425.

\bibitem{dirac}P.A.M.Dirac, Royal Society of London
Proceedings Series A {\bf 268} (1962) 57.


\bibitem{callan}
J.Callan, G.Curtis and J.M.Maldacena, Nucl.Phys. {\bf B513} (1998) 198
\bibitem{gib1}
G.W.Gibbons, Nucl.Phys.
{\bf B514} (1998) 603
\bibitem{gib2}
G.W.Gibbons, Rev.Mex.Fis.
{\bf 49S1} (2003) 19
\bibitem{sen}
A.Sen, JHEP
{\bf 04} (2002) 048


\bibitem{scherrer}
R.J. Scherrer ,Phys.Rev.Lett.
{\bf 93} 011301 (2004)

\bibitem{chimento1}
L.P.Chimento, Phys.Rev.{\bf D69} 123517 (2004) [astro-ph/0311613].

\bibitem{odin1}
E.~Elizalde, S.~Nojiri, S.~D.~Odintsov, D.~Saez-Gomez, Eur.Phys.J.C70,351-361 (2010)[aRxiv:1006.3387]

\bibitem{odin2}
S.~Nojiri,S.~D.~Odintsov
[arXiv:1008.4275]

\bibitem{odin3}
E.~Elizalde, S.~Nojiri, S.~D.~Odintsov, L.~Sebastian, S.~Zerbini, Phys.Rev.{\bf D83} 086006 (2011); [arXiv:1012.2280] 

\bibitem{simon} 
  J.~Simon, L.~Verde and R.~Jimenez,
  Phys.\ Rev.\ D {\bf 71}, 123001 (2005)
 
 \bibitem{barger}
  V.~D.~Barger and D.~Marfatia,
  Phys.\ Lett.\  B {\bf 498}, 67 (2001)
  [arXiv:astro-ph/0009256].

\bibitem{aasen}
A.~A.~Sen, JCAP {\bf 0603} , 010 (2006)


\bibitem{ref:tsujikawa}
  S.~Tsujikawa,
  arXiv:hep-ph/0304257.

\bibitem{ref:mukhanov}
 V.~Mukhanov, ``Physical Foundations of Cosmology", Cambridge (2005)
 \bibitem{ref:paddy} T.~Padmanabhan and T.~R.~Choudhury,
  Mon.\ Not.\ Roy.\ Astron.\ Soc.\  {\bf 344}, 823 (2003)

 \bibitem{Davis} T.~ M.~Davis {\it et al.}, 
  Astrophys. J. {\bf 666}, 716 (2007)
   
 \bibitem{Kessler} R.~Kessler  {\it et al.},  
  Astrophys.\ J.\ Suppl.\  {\bf 185 }, 32 (2009)

 \bibitem{Amanullah}  R.~Amanullah {\it et al.}, 
  Astrophys.\ J.\  {\bf 716}, 712 (2010)
  
  \bibitem{Xu}  L.~Xu, Y.~Wang,   JCAP {\bf 1006}, 002 (2010)

  \bibitem{Abraham}  R.~G.~Abraham {\it et al.},
  Astron.\ J.\  {\bf 127}, 2455 (2004)
  
  
  \bibitem{x39} E.~Gaztanaga, A.~Cabre, L.~Hui,
 Mon.\ Not.\ Roy.\ Astron.\ Soc. {\bf 399}, 166 (2009)
 
 \bibitem{x40}
A.~G.~Riess  {\it et al.}, Astrophys. J.  {\bf 699},  539 (2009)

\bibitem{x41}
D.~Stern, R.~Jimenez, L.~Verde, M.~Kamionkauski, S.~A.~.Stanford,
 JCAP {\bf 1002}, 008 (2010)

\bibitem{gango1}
D.Gangopadhyay and S. Mukherjee,
 Phys. Lett.{\bf B665} 121 (2008)
 
\bibitem{gango2}
D.Gangopadhyay, Gravitation and Cosmology
{\bf 16} 231 (2010)

 
 

\end{thebibliography}
\end{document}